\documentstyle[twoside,fleqn,epsfig]{article}
\def\Journal#1#2#3#4{{#1} {\bf #2} (#3) #4}

\def\NCA{\em Nuovo Cimento A}

\def\PLB{{\em Phys. Lett.}  B}
\def\PRL{\em Phys. Rev. Lett.}
\def\PRD{{\em Phys. Rev.} D}

\def\APP{{\em Astrop. Phys.}}

\def\be{\begin{equation}}
\def\ee{\end{equation}}
\def\bea{\begin{eqnarray}}
\def\eea{\end{eqnarray}}
\newcommand{\lsim}{\mathrel{\mathop{\kern 0pt \rlap
  {\raise.2ex\hbox{$<$}}}
  \lower.9ex\hbox{\kern-.190em $\sim$}}}
\newcommand{\gsim}{\mathrel{\mathop{\kern 0pt \rlap
  {\raise.2ex\hbox{$>$}}}
  \lower.9ex\hbox{\kern-.190em $\sim$}}}

\newcommand{\AmS}{{\protect\the\textfont2
  A\kern-.1667em\lower.5ex\hbox{M}\kern-.125emS}}
\normalsize
\begin{document}

\begin{flushright}
{\bf ROM2F/2006/08\\}
{\bf to appear on Eur. Phys. J. C \\}
\end{flushright}

\baselineskip=0.65cm

\begin{center}
\Large 
{\bf Investigating halo substructures with annual modulation signature } \\
\rm
\end{center}

\vspace{0.5cm}
\normalsize

\noindent \rm R.\,Bernabei,~P.\,Belli,~F.\,Montecchia,~F.\,Nozzoli

\noindent {\it Dip. di Fisica, Universita' di Roma "Tor Vergata"
and INFN, sez. Roma2, I-00133 Rome, Italy}

\vspace{3mm}

\noindent \rm F.\,Cappella, A.\,Incicchitti,~D.\,Prosperi

\noindent {\it Dip. di Fisica, Universita' di Roma "La Sapienza"
and INFN, sez. Roma, I-00185 Rome, Italy}

\vspace{3mm}

\noindent \rm R.\,Cerulli

\noindent {\it Laboratori Nazionali del Gran Sasso, I.N.F.N., Assergi,
Italy}

\vspace{3mm}

\noindent \rm
C.J.\,Dai,~H.L.\,He,~H.H.\,Kuang,~J.M.\,Ma,~X.D.\,Sheng,~Z.P.\,Ye\footnote{also
University of Jing Gangshan, Jiangxi, China}

\noindent {\it IHEP, Chinese Academy, P.O. Box 918/3, Beijing 100039,
China}

\vspace{3mm}

\noindent \rm M.\,Martinez

\noindent {\it Laboratory of Nuclear and High Energy Physics, University of 
Zaragoza, 50009 Zaragoza, Spain}

\vspace{3mm}

\noindent \rm G.\,Giuffrida

\noindent {\it Dip. di Fisica, Universita' di Roma "Tor Vergata", I-00133 Rome, Italy
and INAF, Osservatorio Astronomico di Roma, I-00040 Monte Porzio Catone, Italy}

\normalsize

\section{Abstract}
Galaxy hierarchical formation theories, numerical simulations, the                                    
discovery of the Sagittarius Dwarf Elliptical Galaxy (SagDEG) in 1994 
and more recent investigations 
suggest that the dark halo of the Milky Way                   
can have a rich phenomenology containing non thermalized
substructures. In the present preliminary study, we investigate 
the case of the SagDEG (the best known satellite galaxy in the Milky Way 
crossing the solar neighbourhood) 
analyzing the consequences of its dark matter stream contribution to the 
galactic halo on the basis of the DAMA/NaI annual modulation data. 
The present analysis is restricted to some WIMP candidates and
to some of the astrophysical, nuclear and 
particle Physics scenarios.
Other candidates such as e.g. the light bosonic ones, we discussed
elsewhere, and other non thermalized 
substructures are not yet addressed here.

{\it Keywords:} Dark Matter, Sagittarius Dwarf Elliptical Galaxy

{\it PACS numbers:} 95.35.+d

\section{Introduction}
 
The DAMA/NaI set-up \cite{NaI99,Sist,RNC,IJMPD} has exploited the model independent annual modulation 
signature over seven annual cycles \cite{NaI99,Sist,RNC,IJMPD,Mod1,Mod2,Ext,Mod3,Sisd,Inel,Hep,IJMPA},
achieving 6.3 $\sigma$ C.L. model independent 
evidence for the presence of a Dark Matter (DM) particle component in the galactic halo. 
Some of the many possible corollary quests for the candidate particle have been carried out 
both on the WIMP class of DM candidate particles with various features and increasing
exposures  
and on keV-range pseudoscalar and scalar DM candidate particles (to which 
experimental activities, applying whatever rejection technique of the electromagnetic component of the
counting rate, are blind). Various possibilities 
for the candidate and the interactions have also been discussed in literature by 
e.g. \cite{Bo03,Bo04,Botdm,khlopov,Wei01,foot,Saib}.

Many of the uncertainties and assumptions affecting whatever kind of model dependent result 
in the field 
(such as e.g. corollary quests for candidate, 
exclusion plots, and --in the case of indirect investigation experiments --
the determination of evidence itself, of the parameters or of the limits, etc.)
have been discussed in some details e.g. in ref. \cite{RNC}.

Here we will make a preliminary investigation on the effect of DM stream contributions in
the galactic halo restricting the presentation to some solutions for the WIMP class 
already discussed e.g. in refs. \cite{RNC,IJMPD} and -- as regards the stream contribution 
-- to the SagDEG case,
which has been already addressed in a different way in literature e.g. in ref. \cite{Fre04,Gelmini}. 
Other interesting cases and candidates will be further addressed in this 
light in future works. 

As known, DAMA/NaI exploited the effect of the Earth revolution 
around the Sun on the DM particles' interactions in the target-material
of suitable underground detectors. 
As a consequence of its annual revolution, the Earth should be 
crossed by a variable flux of DM particles along the year. 
In particular, the expected differential rate
as a function of the recoil energy, $dR/dE_R$ (see ref. \cite{RNC} for detailed 
discussion), depends on time owing to 
the DM particle velocity distribution in the
laboratory frame, $f(\vec{v}|\vec{v}_{\oplus}(t))$; here
$\vec{v}_{\oplus}(t)$ is the Earth's velocity in the galactic frame
as function of time.

This method offers an efficient model independent signature, able to test a large interval of 
cross sections and of halo densities; it is named {\it annual modulation 
signature} and was originally suggested in the middle of '80 by \cite{Freese}. 

In particular the expected counting 
rate averaged in a given energy interval
can be expressed by the first order Taylor approximation:
$S(t) \simeq S_0 + S_m cos\omega(t-t_0)$
with the contribution from the highest order terms less than 0.1$\%$.
$S_0$ is the unmodulated term, $S_m$ is the modulation amplitude,
$\omega$= 2$\pi$/T with T=1 year and $t_0$, time when the 
expected counting rate is maximum, depends on the adopted halo model
and on possible non thermalized contributions.  
In particular, for halo models with velocity distribution
isotropic in the galactic frame $t_0$ is roughly June 2$^{nd}$ 
when the Earth velocity in the galactic frame is at maximum.
In the present paper we use the following prescription
for the velocity distribution in the laboratory frame once fixed 
the halo model:
$
\rho_{tot} \times f(\vec{v}|\vec{v}_{\oplus}(t))
=
\rho_{halo} \times f_{halo}(\vec{v}|\vec{v}_{\oplus}(t))
+
\rho_{stream} \times f_{stream}(\vec{v}|\vec{v}_{\oplus}(t))$, 
where the two contributions of the DM particles in the 
dark halo and the DM particles in the stream have been
pointed out. Here $\rho_{tot}$, 
$\rho_{halo}$, $\rho_{stream}$ are the DM particle
densities and $f_{halo}$ and $f_{stream}$ are the 
velocity distributions of the two components normalized to one.

\section{SagDEG phenomenology}

Since the discovery of the SagDEG in 1994
\cite{Ibata}, it has been argued that the dark halo of the Milky Way 
can have a rich phenomenology containing non thermalized substructures. This 
hypothesis is also supported by galaxy hierarchical formation 
theories \cite{Navarro96} and by some numerical simulations \cite{moore}.
Additional interest is offered by the observation of other satellites of the Milky Way, 
such us the Canis Major in 2003 \cite{bellaz}, and satellites of other near galaxies, 
like the stream ($v \sim 300$ km/s ) discovered in our nearest neighbouring "twin'' galaxy M31 \cite{chapman}.
In 1998 it was found that the SagDEG orbits the Milky Way Galaxy in about 1 Gy,
having passed through dense central region of our Galaxy at least about 10 times
during its life. This has been interpreted as an indication of presence of DM 
that with its gravity has prevented the disruption of the SagDEG \cite{velasq}.

Suitable DM direct detection experiments can provide interesting
information about the local halo structure, investigating
the presence of non thermalized dark matter fluxes, as in case
of the tidal stream of a dwarf satellite galaxy passing through 
the solar neighbourhood. SagDEG is the best known satellite galaxy of 
the Milky Way crossing the solar neighbourhood; here the consequences of 
its dark matter stream contribution to the galactic halo are 
analyzed on the basis of the DAMA/NaI annual modulation data 
\cite{NaI99,Sist,RNC,IJMPD,Mod1,Mod2,Ext,Mod3,Sisd,Inel,Hep,IJMPA}.
In fact, e.g. -- as reported in ref. \cite{IJMPD} -- the presence of DM streams in the 
Galaxy would induce a (slightly) variation of the phase value of the modulated 
component of the signal and its variation with energy. Consequently, also the 
$\frac{S_m}{S_0}$ ratio undergoes a change depending on the energy window and on the stream properties
\footnote{In particular, the 
$\frac{S_m}{S_0}$ ratio would increase or decrease with the respect to absence of stream in the galactic halo 
depending on the local direction of the stream in the halo.}. 

The SDSS and the 2MASS surveys \cite{SDSS,2MASS} have traced 
the tidal stream of the SagDEG; 
two streams of stars are being tidally pulled away from its main body 
and extend outward from it. The leading tail can shower matter down through 
the solar neighbourhood and considerations, based on the (very uncertain) $M/L$ ratio, 
suggest the allowed density in the SagDEG tail, $\rho_{sgr}$, to be of the order of 
$(0.001-0.07)$GeV~cm$^{-3}$
corresponding to about $(0.3-23)\%$ of the halo local density \cite{Fre04}.

Fundamental information, in order to investigate effects correlated with the presence of
such tidal streams, is the value of the mean velocity of the stream, its direction and its velocity
dispersion.
Despite the fact that SagDEG is the best known satellite galaxy of the Milky Way 
crossing the solar neighbourhood, these quantities are not yet well
defined and a large number of related investigations can be found in literature.
In particular, in this paper we have taken into account both the values
of ref. \cite{Fre04} (derived from the analysis of eight clump stars
-- from Chiba and Yoshii catalogue \cite{chiba} -- attributed to the SagDEG tail) and
the values of ref. \cite{Law} (based on a SagDEG simulation model).

In the following, we will use a right-handed reference frame with the $x$ 
axis towards the Galactic Center,
the $y$ axis towards the direction of Galaxy rotation and the $z$ axis 
towards the Galactic North Pole. 

To account for the determination of ref. \cite{Fre04}, the SagDEG stream has been modeled 
as a DM flux with mean velocity in galactic coordinates given by:
\begin{equation}
\vec{V}_{8*}=(V_x, V_y, V_z) = (-65\pm22,135\pm12, -249\pm6)~km/s.
\label{eq:velSag1}
\end{equation}

\begin{figure}[ht]
\centering
\includegraphics[width=120pt] {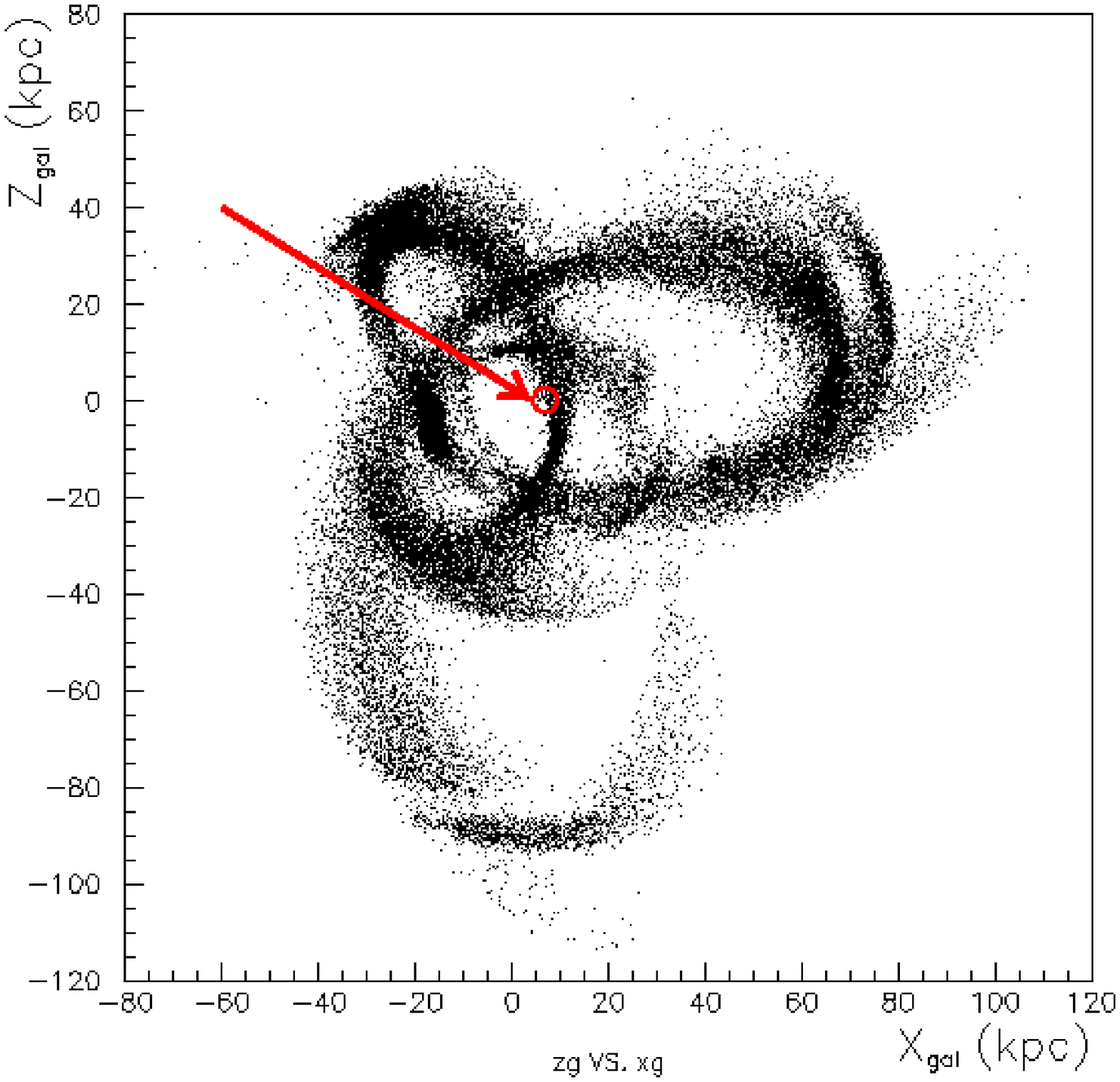}
\includegraphics[width=120pt] {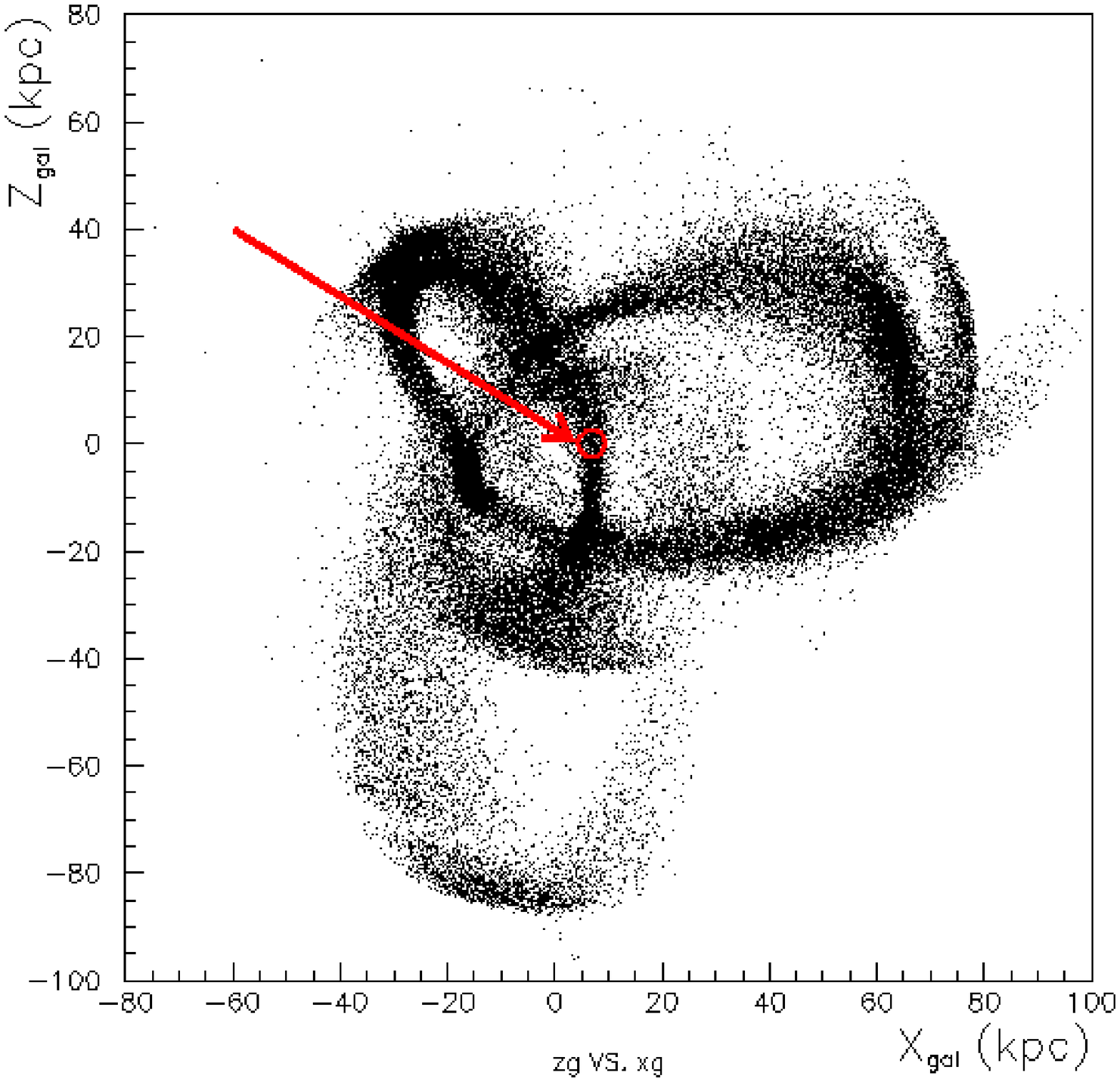}
\includegraphics[width=120pt] {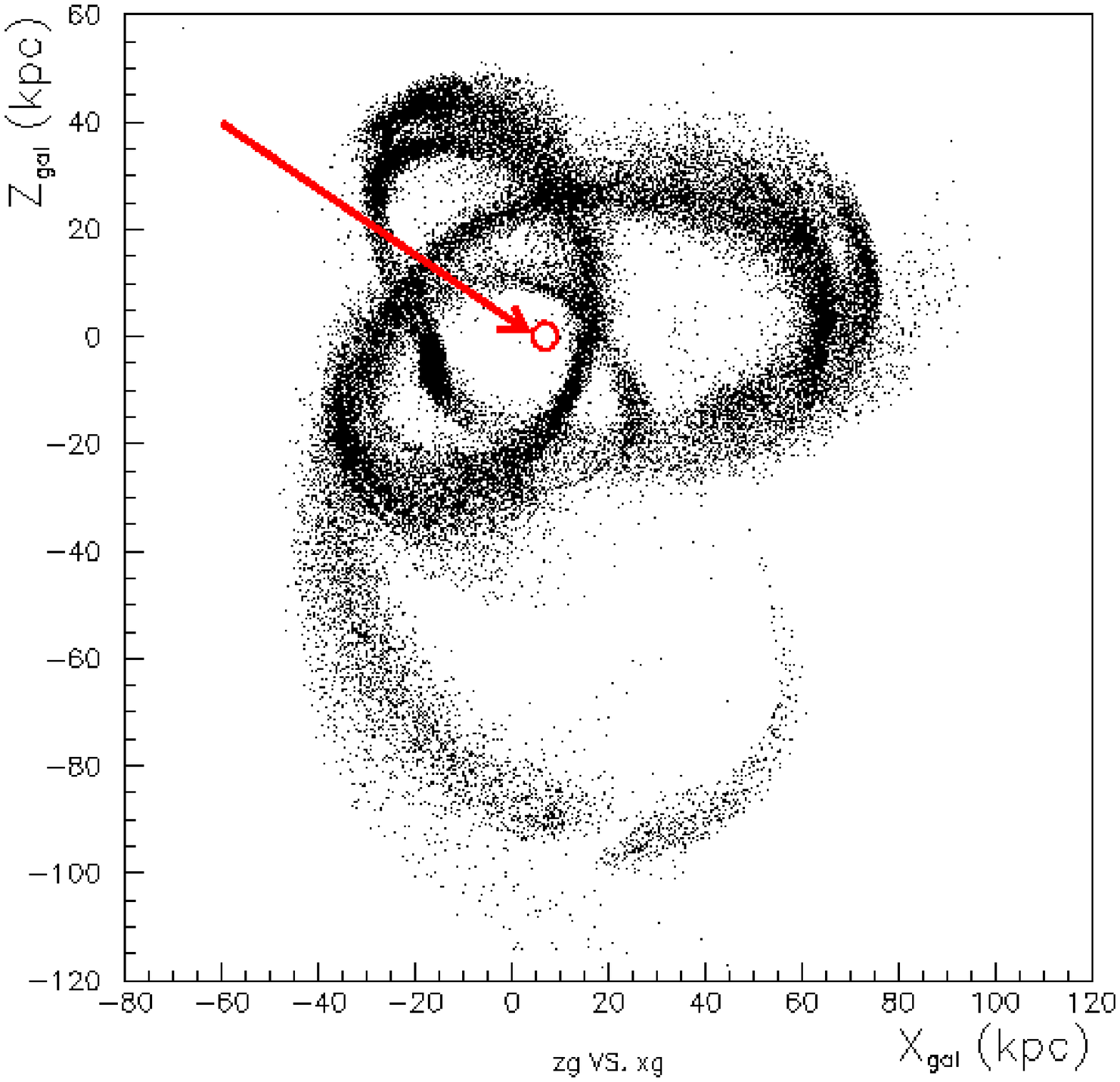}
\caption{SagDEG simulation models 
for spherical (left), oblate (center) and prolate (right)
halo potentials; data taken from ref. \cite{Law}. 
In each panel the circle pointed by the arrow selects the Earth position and the 
configurations considered in this paper for the evaluation of the used mean 
velocity values: $\vec{V}_{sph}$ and $\vec{V}_{obl}$.
We note that no configuration is present in the solar neighbourhood for the prolate model.}
\label{fg:models}
\end{figure}

\begin{figure}[ht]
\centering
\includegraphics[width=250pt] {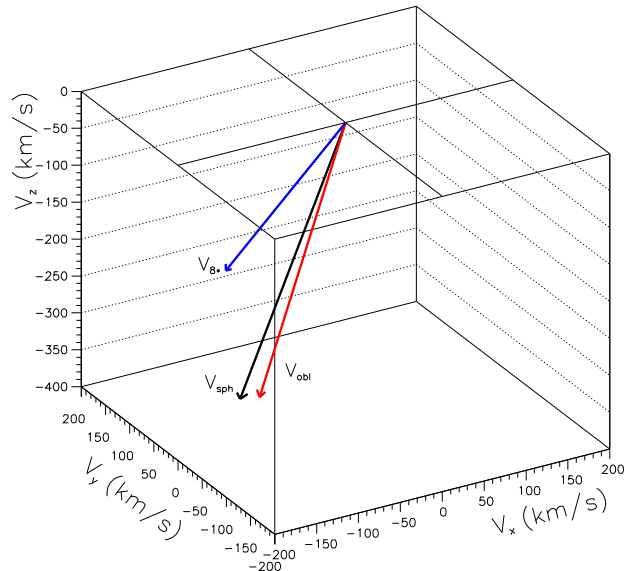}
\vspace{-0.8cm}
\caption{Graphical representation of the three stream mean velocity sets studied in the
present paper for the SagDEG tidal stream. }
\label{fg:vel}
\end{figure}

\noindent Here 1 $\sigma$ error has been reported for each velocity component. 
In addition, to account for the determination of ref. \cite{Law} we
have also considered the following cases :

\begin{equation}
\vec{V}_{sph}=(V_x, V_y, V_z) = (-86\pm14,69\pm3,-384\pm1)~km/s ,
\label{eq:velSag2}
\end{equation}

\noindent and
\begin{equation}
\vec{V}_{obl}=(V_x, V_y, V_z) = (-57\pm8,79\pm3, -395\pm1)~km/s ,
\label{eq:velSag3}
\end{equation}

\noindent for the spherical and the oblate halo models of ref. \cite{Law}, respectively.
These two last stream mean velocities 
have been derived by
considering for each halo model the $\simeq$ 100 configurations nearest
to the Sun within a distance $\lsim$ 2.5 kpc (see Fig. \ref{fg:models}).
It is worth to note that
the prolate model of ref. \cite{Law} 
has not been considered in the following 
since no configuration is present in the solar neighbourhood. A graphical representation 
of the three stream mean velocity sets, considered in the
present paper for the SagDEG tidal stream, is shown in Fig. \ref{fg:vel}.

As regards the velocity dispersion to be associated with each one of the three considered stream parametrizations 
given above, we derive for our reference frame, respectively:

\begin{equation}
(\sigma_x, \sigma_y, \sigma_z)_{8*} = (62, 33, 17)~km/s,
\label{eq:velSag1s}
\end{equation}

\begin{equation}
(\sigma_x, \sigma_y, \sigma_z)_{sph} = (60, 19, 8)~km/s 
\label{eq:velSag2s}
\end{equation}

and

\begin{equation}
(\sigma_x, \sigma_y, \sigma_z)_{obl} = (59, 23, 9)~km/s.
\label{eq:velSag3s}
\end{equation}

\noindent The $(\sigma_x, \sigma_y, \sigma_z)_{8*}$ is taken from ref.~\cite{Fre04}, while 
$(\sigma_x, \sigma_y, \sigma_z)_{sph}$ and $(\sigma_x, \sigma_y, \sigma_z)_{obl}$ have been calculated 
for each model as r.m.s.~values of the about 100 configurations in the solar neighbourhood.
As it can be observed, for the three considered SagDEG stream models, the velocity dispersions are quite
different and significantly non-isotropic.
Notwithstanding,
in the following for simplicity the velocity distribution 
of the SagDEG stream in the solar neighbourhood has been approximated by an isotropic 
Maxwellian distribution in the locally comoving frame of SagDEG.
The investigation of the effect of non-isotropic distributions is not considered in the present
paper and can be addressed in future.

\subsection{A full example of the SagDEG effect on the annual modulation signature for a given scenario}

In this subsection we show for template purpose a complete example of the 
effect induced by the presence of a DM stream in the solar neighbourhood considering 
a particular scenario.

As known, the Earth velocity in galactic coordinates can be expressed 
as following:
\begin{equation}
\vec{v}_{\oplus}(t) =\vec{v}_{LSR} + \vec{v}_{\odot} + 
  V_{Earth}(\hat{e}_1\sin\lambda(t) - \hat{e}_2\cos\lambda(t))
\label{eq:velearth}
\end{equation}
where $\vec{v}_{LSR}=(0, 220\pm 50, 0)$ km/s
(the quoted uncertainty is at 90$\%$C.L.)  
is the velocity of the Local Standard of Rest; 
$\vec{v}_{\odot}=(10.00, 5.25, 7.17)$ km/s 
is the Sun peculiar velocity here taken from ref. \cite{Dehnen} 
and $V_{Earth}$ is the mean orbital velocity of the Earth ($\simeq$ 29.8 km/s).
The $\hat{e}_1$, $\hat{e}_2$ versors and the $\lambda(t)$ function
are \cite{Gelmini}:
\begin{eqnarray}
\hat{e}_1=(-0.0670, 0.4927, -0.8676), \nonumber \\
\hat{e}_2=(-0.9931, -0.1170, 0.01032), \nonumber \\
\lambda(t)=\omega (t-0.218). \nonumber
\end{eqnarray}

\noindent Here $\omega = 2\pi/T$ with $T$ = 1 y, $t$ is the time in years starting from January 1st  
and 0.218 y is the spring equinox (March 21). 

The velocity distribution of the SagDEG DM particles in the laboratory frame 
can be written as \footnote{We note that in the following, 
the quantities related to SagDEG are marked as $sgr$.}:

\begin{equation}
f_{sgr}(\vec{v}) = \frac{1}{ \pi^{\frac{3}{2}} v_{0,sgr}^3} 
e^{-\frac{(\vec{v}-\vec{v}_{sgr,\oplus})^2}{v_{0,sgr}^2}},
\label{eq:disp}
\end{equation}

\noindent where the mean velocity of the SagDEG DM particles in the 
laboratory frame is: $\vec{v}_{sgr,\oplus}(t)  = \vec{v}_{sgr} - \vec{v}_{\oplus}(t)$; 
$\vec{v}_{sgr}$ will be in turn either 
$\vec{V}_{8*}$ or $\vec{V}_{sph}$ or $\vec{V}_{obl}$. Finally,
for each $\vec{v}_{sgr}$,
the $v_{0,sgr} = \sqrt{\frac{2}{3}}\sigma_{sgr}$ parameter 
is assumed in the following to be either
20 or 40 or 60 km/s.

The  $|\vec{v}_{sgr,\oplus}(t)|$ reaches its maximum value at time $t_{0,sgr}$ defined by:
\begin{equation}
cos \left[\lambda(t_{0,sgr})\right] = \frac{a_2}{\sqrt{a_1^2+a_2^2}}, 
\; \; \; \; \; \;
sin \left[\lambda(t_{0,sgr})\right] = - \frac{a_1}{\sqrt{a_1^2+a_2^2}}, 
\label{eq:sincos}
\end{equation}
with $a_i = \hat{e}_i \cdot (\vec{v}_{sgr} - \vec{v}_{LSR} - \vec{v}_{\odot})$.
Therefore, the mean velocity of the SagDEG DM particles in the laboratory frame
would be maximum around January $10^{th}-14^{th}$,
depending on the considered SagDEG velocity set.
We remind that in absence of SagDEG contribution (that is, $f_{halo}$ 
only contributes to the total DM particles velocity distribution), 
$t_0$ is expected to be roughly 
at the 152.5 day of the year ($\sim$ June 2$^{nd}$). 
Hence, the net effect 
of a SagDEG tail contribution to the local halo density is  
a shift of few days (towards January) in
the expected phase of the signal. 
For the sake of completeness, 
the $\frac{S_m}{S_0}$ ratio is not expected to be enhanched 
when a SagDEG stream is included, due to the nearly opposite 
phases between the two extreme cases.

\begin{figure}[ht]
\centering
\includegraphics[width=300pt] {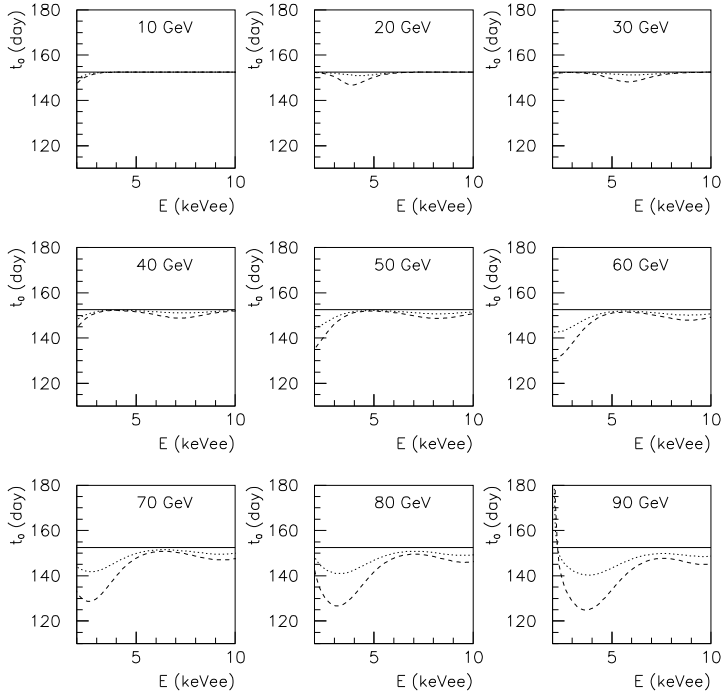}
\caption{Examples of the effect of the SagDEG tail, modeled as given in the text,
on the expected annual modulation signature in NaI(Tl) detectors.
In each panel a plot of the phase ($t_0$) vs the detected energy ($E$) 
is shown for a given WIMP mass and for the given assumptions
(see text).
Dotted line: $\rho_{halo}=0.74$ GeV cm$^{-3}$ (maximum value allowed for the adopted halo
model)
\cite{RNC}; 
dashed line: $\rho_{halo}=0.33$ GeV cm$^{-3}$ (minimum value allowed for the adopted halo model)
\cite{RNC};
solid line: absence of SagDEG contribution, that is $t_0 \sim$ June 2$^{nd}$. 
The effect of a possible SagDEG contribution is to 
slightly shift the phase $t_0$ towards lower values at low
recoil energy.}
\label{fg:nfw_t0}
\end{figure}

As examples, in Fig. \ref{fg:nfw_t0} the phase, $t_0$, of the modulated component of the signal 
is plotted -- for some models with the inclusion of the SagDEG stream 
and for some reference WIMP masses --
as a function of the detected energy in NaI(Tl) detectors.
In particular, in the given examples, simple assumptions have been adopted:

\vspace{0.2cm}
i) $\vec{v}_{sgr} = (-65,135,-249)$ km/s, that is $\vec{V}_{8*}$ at its central value;

\vspace{0.1cm}
ii) $v_{0,sgr} = 40 $  km/s;

\vspace{0.1cm}
iii) $\vec{v}_{LSR}=(0, 220, 0)$ km/s , that is at its central value;

\vspace{0.1cm}
iv) SagDEG tail DM density $\rho_{sgr}=0.04 \times \rho_{halo}$;

\vspace{0.1cm}
v) Galactic halo model: NFW ($\alpha = 1$, $\beta = 3$, $\gamma = 1$, $a = 20$ kpc) (A5 
of \cite{RNC});

\vspace{0.1cm}
vi) WIMP DM candidate with dominant Spin Independent coupling ($\sigma \propto A^2$);

\vspace{0.1cm}
vii) Form factors and quenching factors of $^{23}$Na and $^{127}$I 
as in case A of ref. \cite{RNC}; 
that is, the most cautious Helm form factor, the mean nominal values 
for the parameters of the nuclear form factors and for the measured 
$^{23}$Na and $^{127}$I quenching factors are assumed.

For the sake of completeness, we remind that the DAMA/NaI results (107731 kg$\cdot$day exposure) 
provide $t_0$ = $(140 \pm 22)$ day averaged in the (2-6) keV energy window; at present level of 
sensitivity -- as it can be seen in Fig.~\ref{fg:nfw_t0} and extensively
in the following sections --
it is consistent both with presence and with absence of SagDEG contribution.
As discussed in \cite{IJMPD} larger exposures, which will be available in near 
future thanks to the presently running DAMA/LIBRA set-up \cite{vulcano},
will offer the possibility of more stringent constraints.

\subsection{Investigating the effect of a SagDEG contribution for
WIMP cases}

In order to further investigate the effect of the presence of a SagDEG
stream,  we will follow in this section the same approach already exploited 
in ref. \cite{RNC,IJMPD}, where the simplier case without SagDEG contribution
was considered.
In particular we have considered here 
the WIMP class of candidate particles in the general case of 
mixed SI(Spin Independent)\&SD(Spin Dependent) coupling and the two subcases 
of pure SI and  pure SD couplings. 

\begin{figure}[!p]
\centering
\includegraphics[width=400pt] {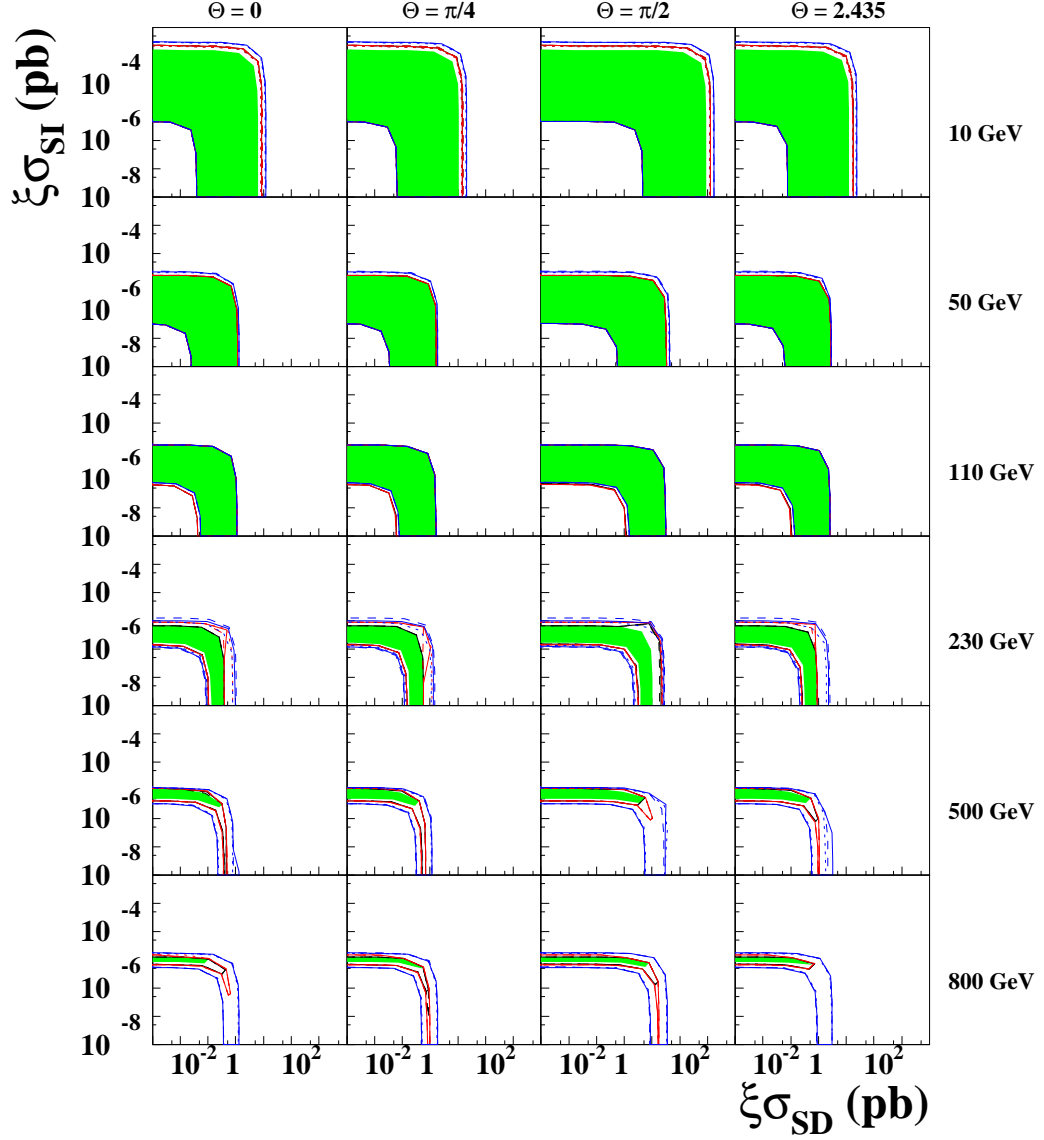}
\vspace{-0.8cm}
\caption{Examples of slices of the four-dimensional allowed volume in the 
($\xi\sigma_{SI},\xi\sigma_{SD}$) plane for some $m_W$ and $\theta$ values
in the considered scenarios.
The shaded regions have been determined for no SagDEG contribution \cite{RNC}, 
while the areas enclosed by the lines are obtained by introducing in the analysis 
the SagDEG stream with DM density not larger than 0.1 GeV cm$^{-3}$. 
The nine  considered possibilities for the SagDEG stream veloctity 
($\vec{V}_{8*}$ (blue), $\vec{V}_{sph}$ (black), $\vec{V}_{obl}$ (red))
and $v_{0,sgr}$ dispersion (20 km/s (dashed),
40 km/s (solid) and 60 km/s (dotted)) have been reported.}
\label{fg:sisd}
\end{figure}

In fact, since the $^{23}$Na and $^{127}$I are fully sensitive to both SI and SD
interactions the most general case is given by a four-dimensional volume 
($m_W$, $\xi\sigma_{SI}$, $\xi\sigma_{SD}$, $\theta$), where $m_W$ is the DM particle mass,
$\xi$ is the ratio between the local density for the considered candidate and 
the local DM density $\rho_{tot}$, $\sigma_{SI}$ is the SI WIMP-nucleon cross section and 
$\sigma_{SD}$ is the SD WIMP-nucleon cross section according to the definitions and 
scaling laws considered in ref. \cite{RNC}; $tg\theta$ is the ratio between
the effective coupling strengths to neutron and proton for the SD couplings ($\theta$ can vary between 0 and $\pi$).
In the calculation the same galactic halo models and associated parameters as in ref. \cite{RNC} have been considered
as well as the uncertainty on the value of the local velocity $v_0 = (220\pm50)$ km/s (90\% C.L.).
For the case of the SagDEG stream description we have  
considered the three possibilities for the velocity ($\vec{V}_{8*}$, $\vec{V}_{sph}$ and $\vec{V}_{obl}$
at their central values) and three possible velocity dispersions ($v_{0,sgr}=20,40,60$ km/s),
obtaining nine different cases for each fixed halo model. 
Moreover, in this section we consider that the SagDEG contribution cannot exceed 
$\sim$0.1 GeV cm$^{-3}$,
as suggested by M/L ratio considerations of ref.~\cite{Fre04}.

The results presented by DAMA/NaI on the corollary quests for the WIMP candidate particles
over the seven annual cycles are calculated here and elsewhere (e.g. \cite{RNC,IJMPD} and references 
therein) taking into
account the time and energy behaviours of the {\it single-hit} experimental data.
For this purpose, the likelihood function 
$L^{i_m}_{i_s,\rho_{sgr}}(m_W,\xi\sigma_{SI},\xi\sigma_{SD},\theta)$ is constructed 
for any fixed SagDEG velocity set and velocity dispersion (cumulatively labeled here as $i_s$) and for
all the considered model frameworks 
(cumulatively labeled here as $i_m$, running on the galactic halo models and 
on all the other parameters involved in the calculation).
In particular, the likelihood function 
requires the agreement: i) of the expectations for the modulated part of the signal 
with the measured modulated behaviour for each detector and for each energy bin; ii) 
of the expectations for the unmodulated component of the signal with the respect 
to the measured differential energy distribution; iii) for WIMP candidate (since ref. \cite{Mod3})
also with the 
bound on recoils obtained by pulse shape discrimination 
from the devoted DAMA/NaI-0 data \cite{Psd96}.
The latter one (used when WIMP candidates are considered)
acts in the likelihood procedure as an experimental upper bound on the 
unmodulated component of the 
signal and -- as a matter of fact -- 
as an experimental lower bound on the estimate of the background levels by the maximum
likelihood procedure.
Thus, the C.L.'s, we quote for  
allowed regions, already account for compatibility with the measured differential
energy spectrum and, -- for WIMP candidates -- with the measured upper bound on recoils.
In particular, in the following for simplicity, the results of these corollary quests for the
candidate particle are presented in terms of allowed regions
obtained as superposition of the configurations corresponding
to likelihood function values {\it distant} more than $4\sigma$ from
the null hypothesis (absence of modulation) in each of the several 
(but still a limited number) of the possible 
model frameworks considered here. 
Obviously, these results are not exhaustive of the many scenarios
possible at present level of knowledge (e.g. for some other recent ideas 
see \cite{Kam03,Sikivie}) and 
larger  
sensitivities than those reported in the following would be reached when including
the effect of other
existing 
uncertainties on assumptions and related parameters \cite{RNC,IJMPD}.

For the general case of a WIMP with mixed SI\&SD coupling, one 
obtains a four-dimensional allowed volume\footnote{Is worth to 
note that -- for example -- experiments 
using either nuclei largely insensitive to SD coupling (as e.g. $^{nat}$Ge, $^{nat}$Si, $^{nat}$Ar, $^{nat}$Ca, 
$^{nat}$W, $^{nat}$O) or nuclei in principle all sensitive to such a coupling but having different 
unpaired nucleon (neutron in odd spin nuclei, such as $^{129}$Xe, $^{131}$Xe, 
$^{125}$Te, $^{73}$Ge, $^{29}$Si, $^{183}$W) with the respect to the proton in $^{23}$Na and $^{127}$I
cannot explore most of the four-dimensional allowed volume.}. 
Since a full picture of this result is not possible in the practice, 
Fig.\ref{fg:sisd} shows some slices of the four-dimensional allowed volume 
in the plane $\xi\sigma_{SI}$ vs $\xi\sigma_{SD}$ for some of the possible $m_W$ and 
$\theta$ values.
The filled areas show the case without SagDEG contribution and, therefore, have already been 
reported and discussed in ref. \cite{RNC,IJMPD} (some different slices are also shown here), while the 
areas enclosed by lines
show the cumulative effect of the possible SagDEG stream
contribution in various cases (see figure caption). 

\begin{figure}[ht]
\centering
\includegraphics[width=200pt] {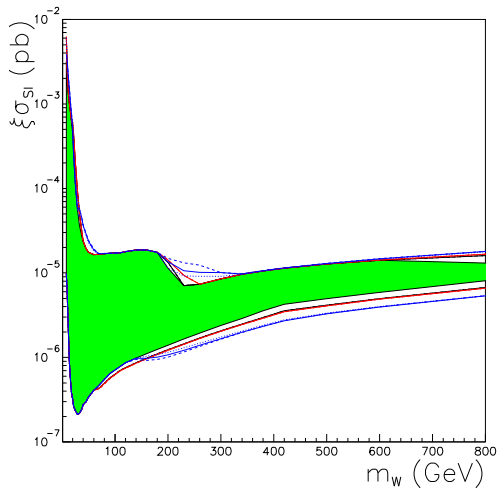}
\vspace{-0.4cm}
\caption{Region allowed in the ($\xi\sigma_{SI},m_W$)
plane in the considered scenarios for pure SI coupling. 
The filled region has been determined for no SagDEG contribution \cite{RNC,IJMPD},
while the areas enclosed by
lines are obtained by introducing 
in the analysis the SagDEG stream with DM density not larger than 0.1 GeV cm$^{-3}$. 
The nine considered possibilities for the SagDEG stream veloctity 
($\vec{V}_{8*}$ (blue), $\vec{V}_{sph}$ (black), $\vec{V}_{obl}$ (red))
and $v_{0,sgr}$ dispersion (20 km/s (dashed),
40 km/s (solid) and 60 km/s (dotted)) have been reported.}
\label{fg:mwsigsi}
\end{figure}
\begin{figure}[ht]
\centering
\includegraphics[width=300pt] {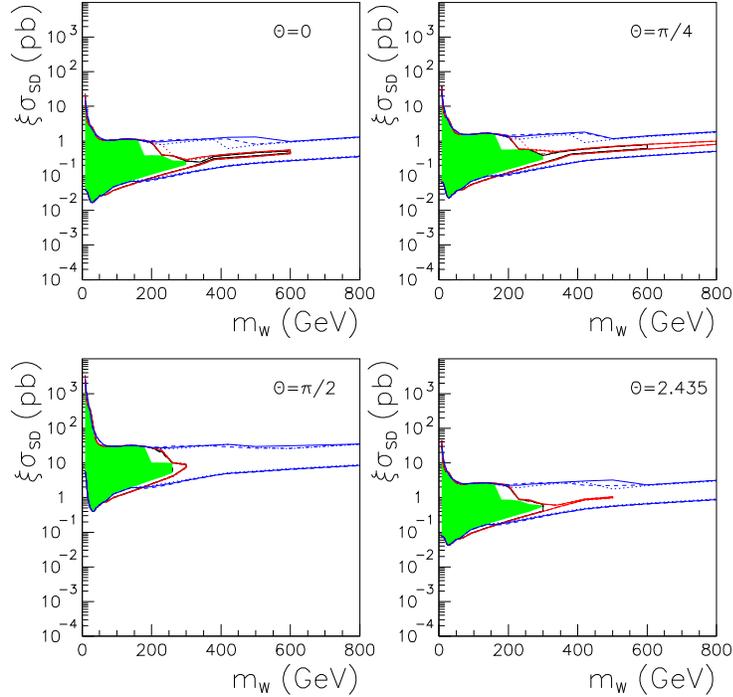}
\vspace{-0.4cm}
\caption{Examples of slices of the three-dimensional allowed volume in the 
($\xi\sigma_{SD},m_W$) plane for some $\theta$ values
in the considered scenarios and for pure SD coupling. 
See Fig. \ref{fg:mwsigsi} for the meaning of the regions.}
\label{fg:mwsigsd}
\end{figure}

The purely SI subcase\footnote{We remind that no direct comparison is possible also among results
on purely SI coupled WIMPs achieved by using different nuclei, although apparently all the presentations 
generally refer to cross section on nucleon. For some discussions on generalities and comparisons see
e.g. \cite{RNC,IJMPD,BOSI}} is shown in Fig.
\ref{fg:mwsigsi},
while in Fig. \ref{fg:mwsigsd}
some slices of the three-dimensional allowed volume ($m_W$, $\xi\sigma_{SD}$, $\theta$) for the purely
SD case are given. 
The filled areas and the areas enclosed by the lines have the same meaning as before.

As it can be observed, the inclusion of the SagDEG stream and of the related uncertainties
significantly modifies the allowed volumes/regions; 
the role appears larger mainly for larger WIMP masses. 

It is worth to note that other streams can potentially play 
more intriguing roles and will be investigated in near future, such as the Canis Major \cite{bellaz}.
Moreover, other kinds of streams as those e.g. arising from caustic halo models \cite{Sikivie} 
can also play a significant role in the corollary investigations for the candidate particle
with whatever approach and for comparisons.

This approach to the problem allows to make some 
cautious constraints on the SagDEG stream in the galactic halo on the basis
of the measured DAMA/NaI annual modulation data. We will discuss some of the implications of the
presented results in the next section.

\section{Constraining the SagDEG stream by DAMA/NaI}

As mentioned, the high exposure of DAMA/NaI can allow to obtain some preliminary information about the 
presence of substructure in the halo as the SagDEG stream.
For this purpose, the likelihood ratio function has been used as statistical analysis approach
in the investigation on the SagDEG parameters with the respect to  
all the others involved in the calculation.

\begin{figure}[!b]
\centering
\vspace{-0.4cm}
\includegraphics[width=170pt] {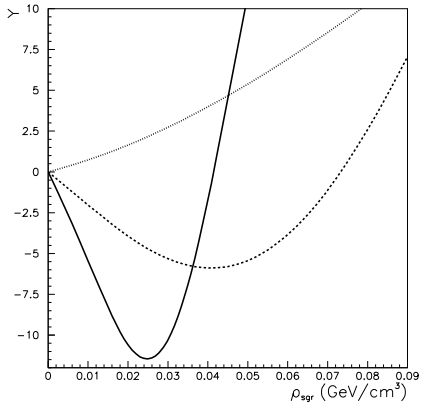}
\vspace{-0.1cm}
\caption{Effect of the SagDEG contribution 
in the data fitting for 3 illustrative models.
The example uses: SI candidate, a SagDEG stream with
velocity set $\vec{V}_{sph}$ and velocity
dispersion $v_{0,sgr} = 40$ km/s; all the parameters for form factors and for the quenching factors are 
fixed at case A of ref. \cite{RNC} (see also sec. 3.1 in the text).
The considered halo models are: i)
NFW halo 
($\alpha = 1$, $\beta = 3$, $\gamma = 1$, $a = 20$ kpc, A5 
of \cite{RNC}),
$v_0 = 220$ km/s, $\rho_{halo}=0.74$ GeV cm$^{-3}$ and $m_W=10$ GeV
(dotted line); 
ii) Evans' logarithmic halo
($R_c = 0$ kpc, $q = 1/\sqrt{2}$, C1 of \cite{RNC}),
$v_0 = 170$ km/s, $\rho_{halo}=0.56$ GeV cm$^{-3}$
and $m_W=22$ GeV (solid line);
iii) Evans' logarithmic counter-rotating halo
($R_c = 5$ kpc, $q = 1/\sqrt{2}$, C2 of \cite{RNC}),
$v_0 = 170$ km/s, $\rho_{halo}=0.67$ GeV cm$^{-3}$,
$\eta = 0.64$ and $m_W=20$ GeV (dashed line).
}
\vspace{-0.3cm}
\label{fg:trovamin}
\end{figure}

In particular, fixing: i) a 
SagDEG velocity set and a velocity dispersion (index $i_s$);
ii) the WIMP mass (labeled as $m_W$) and $\theta$;
iii) the galactic halo
model and all the other parameters involved in the calculation (index $i_m$),
the likelihood ratio as a function of $\rho_{sgr}$ can be defined:
\begin{equation}
\lambda_{m_W}^{i_m,i_s}(\rho_{sgr})
=\frac{max_{\sigma_{(SI,SD)}} \left[ \mathit{L}^{i_m,i_s}_{m_W}(\rho_{sgr}=0) \right]} 
{max_{\sigma_{(SI,SD)}} \left[\mathit{L}^{i_m,i_s}_{m_W}(\rho_{sgr}) \right]},
\label{eq:likeratio}
\end{equation}
\noindent where $max_{\sigma_{(SI,SD)}} \left[\mathit{L}^{i_m,i_s}_{m_W}(\rho_{sgr}) \right]$ 
is the value of the likelihood function
maximized with the respect to the particle cross sections.

\begin{figure}[!b]
\centering
\includegraphics[width=150pt] {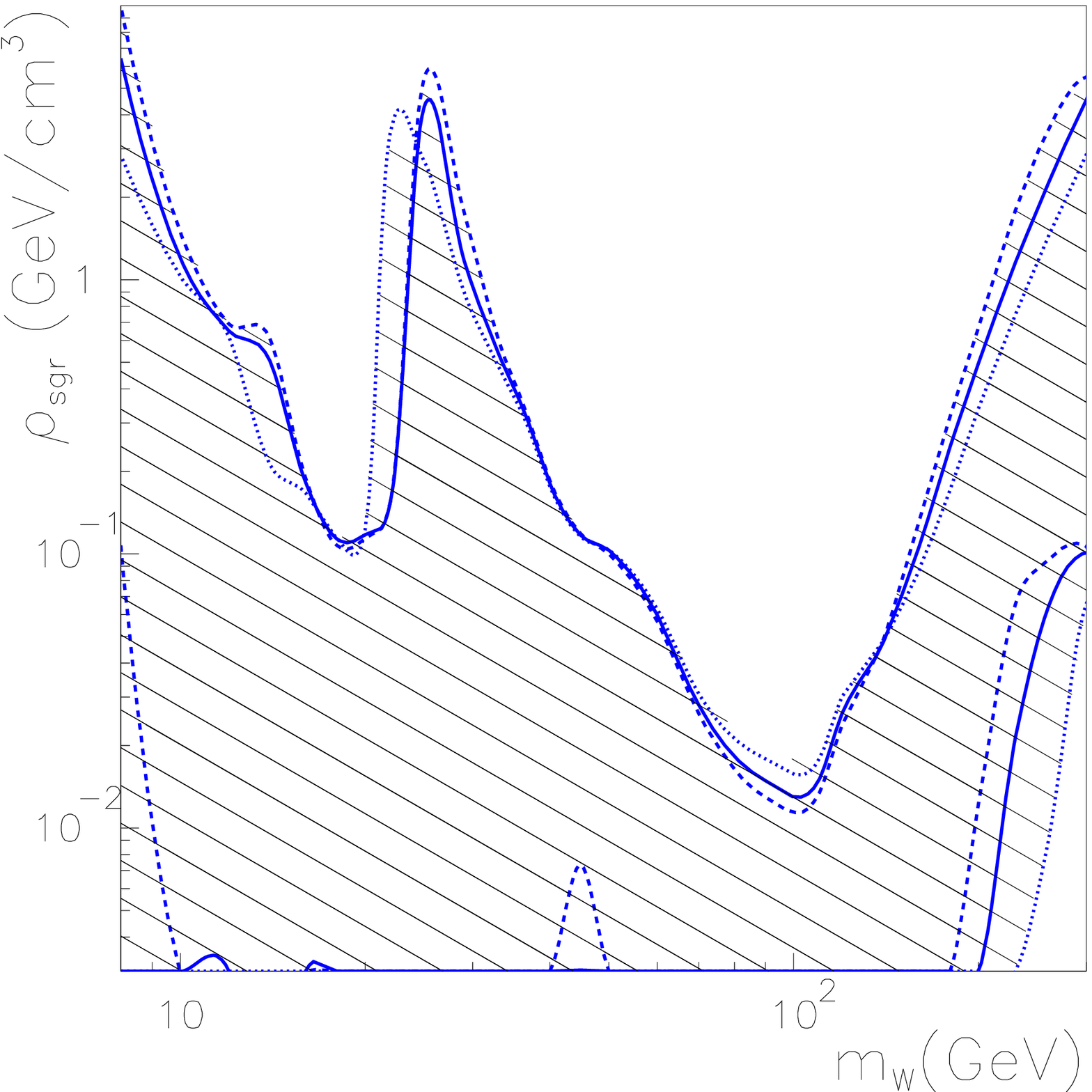}
\includegraphics[width=150pt] {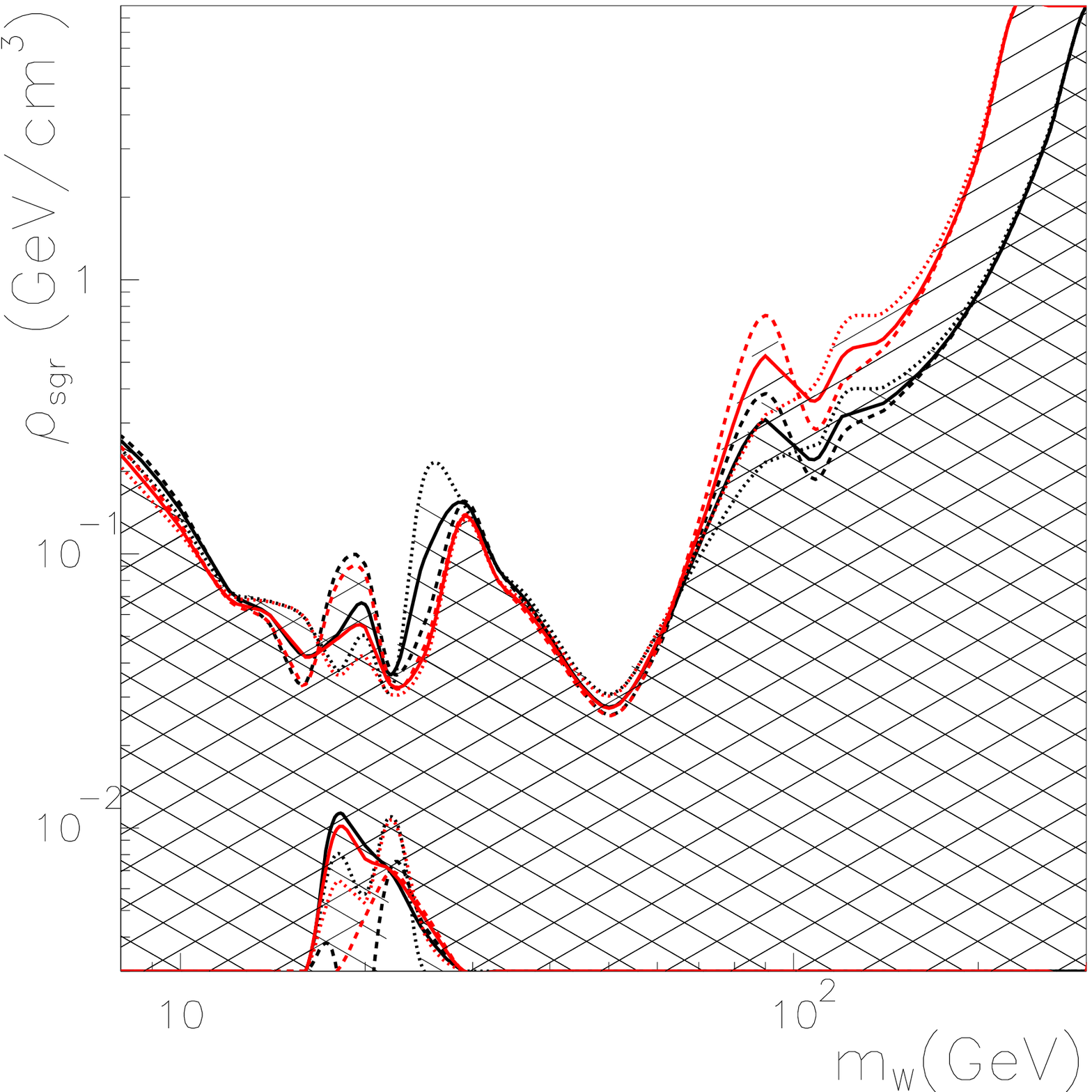}
\caption{
SagDEG density $\rho_{sgr}$ 
allowed at $90\%$ C.L. (hatched area) for pure SI coupling 
as function of $m_W$ values.
Left panel: case of $\vec{V}_{8*}$ velocity set.
Right panel: case of $\vec{V}_{sph}$ (descending hatched) and
$\vec{V}_{obl}$ (ascending hatched) superimposed.
The used stream velocity dispersion, $v_{0,sgr}$ values are: 
20 km/s (dashed), 40 km/s (solid) and 60 km/s (dotted).}
\label{fg:likeratiosi_tot}
\end{figure}

The functions $Y^{i_m,i_s}_{m_W}(\rho_{sgr}) = -2ln(\lambda^{i_m,i_s}_{m_W})$ are 
asymptotically distributed as a chisquare with 1 degree of freedom. 
Some examples
are given in Fig. \ref{fg:trovamin} where
three $Y$ functions are plotted for some halo models and for some particle masses in the
particular case of a pure SI candidate and a SagDEG stream with velocity set 
$\vec{V}_{sph}$ and velocity dispersion $v_{0,sgr} = 40$ km/s.

In particular, this figure shows three representative cases:
i) a model where the SagDEG contribution worsens the data fit
(dotted line);    
ii) a model where the SagDEG contribution improves the data fit
providing a C.L. better than 3 $\sigma$
(solid line);    
iii) a model where the SagDEG contribution improves the data fit
providing a C.L. lower than 3 $\sigma$ (dashed line).

In order to investigate the presence of SagDEG, in all 
the considered halo models and adopted parameter uncertainties,
for simplicity here 
we alternatively investigate only the purely SI and the purely SD cases, respectively.
In the following -- for each considered $m_W$ and $i_s$ -- 
the $90\%$ C.L. allowed intervals on $\rho_{sgr}$ are constructed requiring
that \cite{PDG}: 
\begin{equation}
Y^{i_m,i_s}_{m_W}(\rho_{sgr}) \leq
min_{(i_m,\rho_{sgr})}
\left(
Y^{i_m,i_s}_{m_W}(\rho_{sgr}) \right) +
2.71.
\end{equation}

\begin{figure}[p]
\centering
\includegraphics[width=260pt] {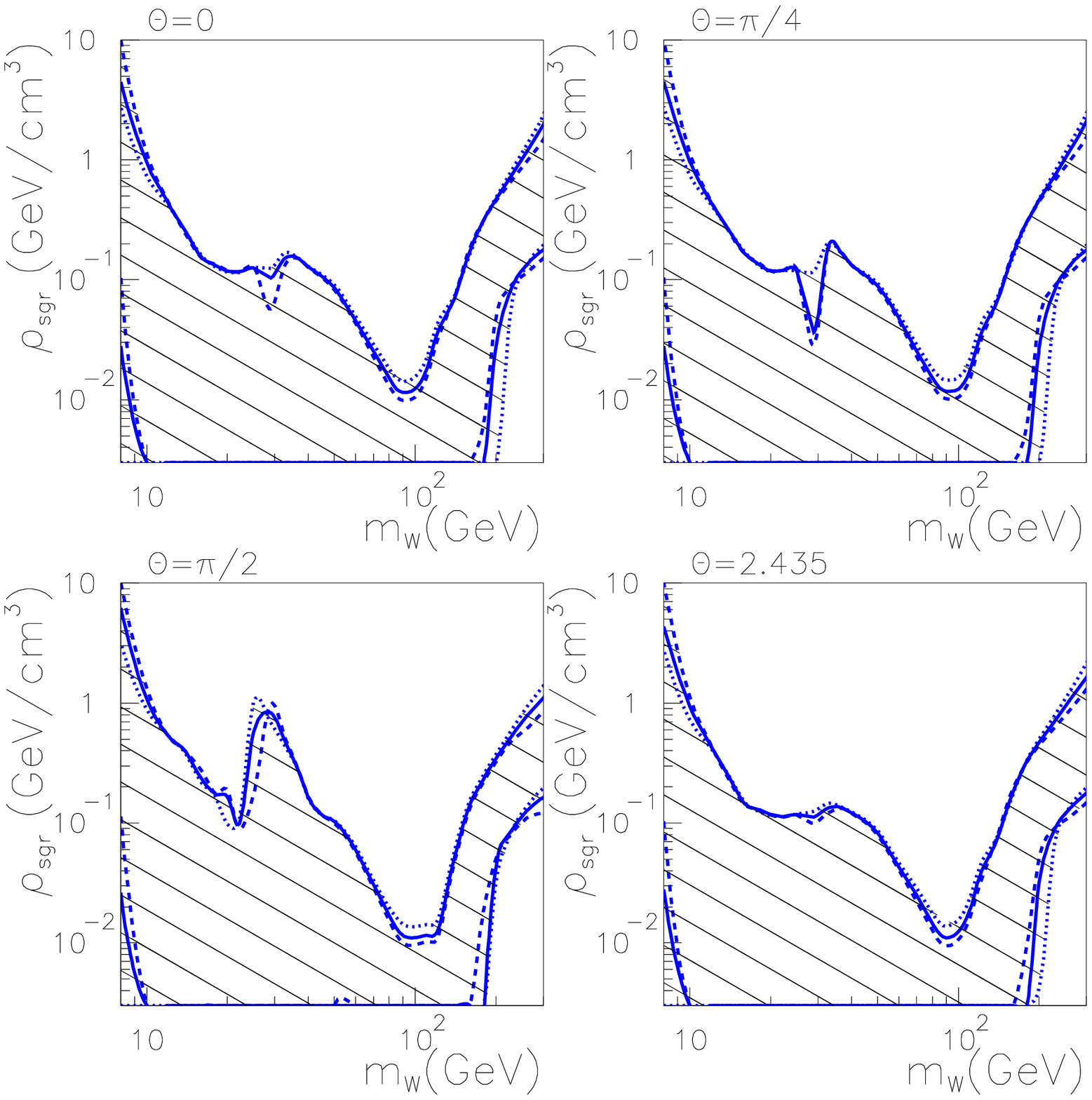}
\includegraphics[width=260pt] {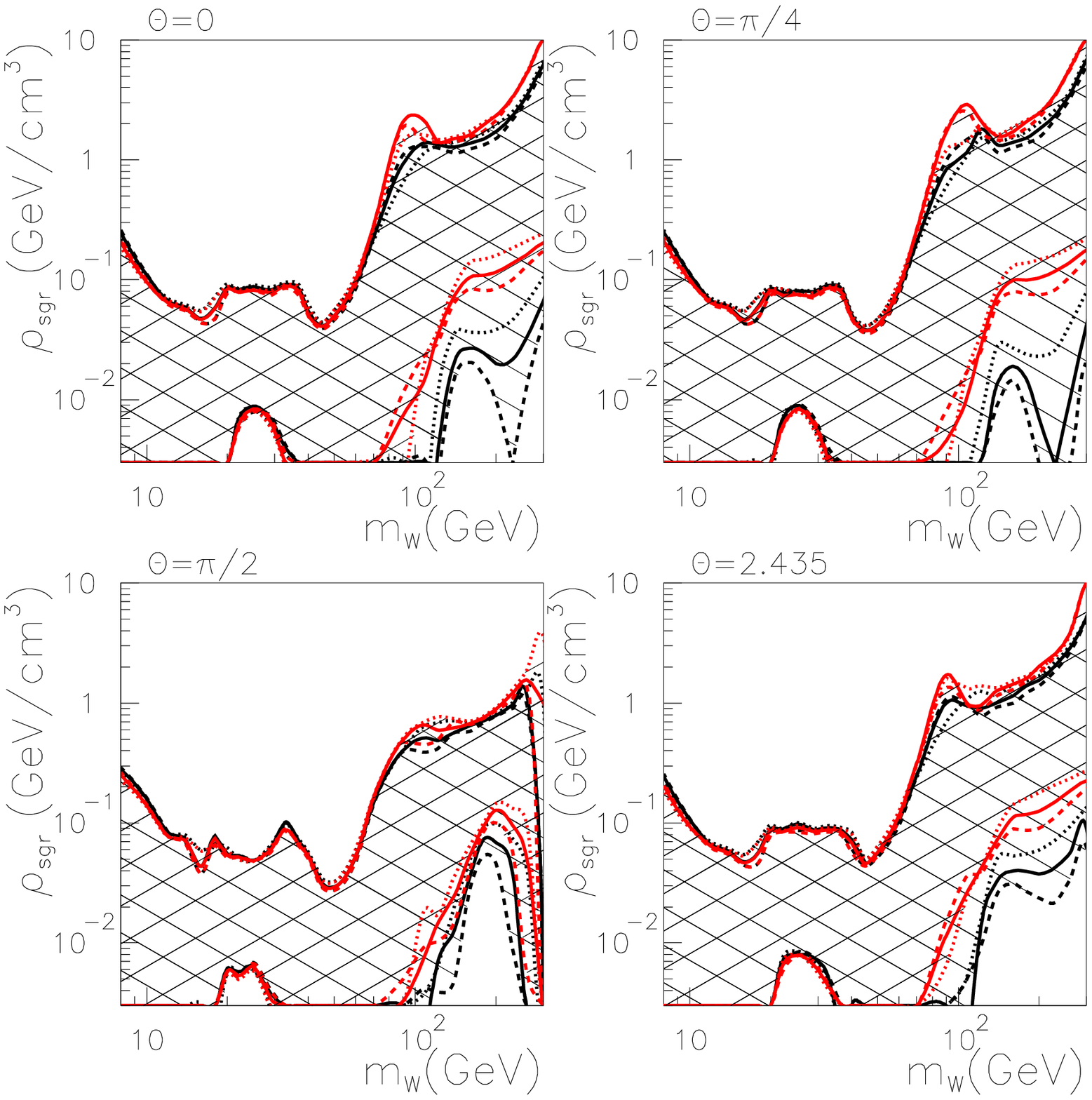}
\vspace{-0.4cm}
\caption{
SagDEG density $\rho_{sgr}$ 
allowed at $90\%$ C.L. (hatched area) for pure SD coupling 
as function of $m_W$ values for some of the possible $\theta$ values.
Upper 4-box: case of $\vec{V}_{8*}$ velocity set.
Lower 4-box: case of $\vec{V}_{sph}$ (descending hatched) and
$\vec{V}_{obl}$ (ascending hatched) superimposed.
The used stream velocity dispersion, $v_{0,sgr}$ values are:
20 km/s (dashed), 40 km/s (solid) and 60 km/s (dotted).}
\label{fg:likeratiosd_tot}
\end{figure}

\noindent In Fig.\ref{fg:likeratiosi_tot} the 
SagDEG density $\rho_{sgr}$
allowed at $90\%$ C.L. 
for pure SI coupling
is shown as a function of $m_W$. 
In Fig. \ref{fg:likeratiosd_tot} the same is shown for a SD coupled candidate
in the particular cases of $\theta=0, \pi/2, \pi/4$ and 2.435 (pure $Z_0$ coupling).
 
From the figures \ref{fg:likeratiosi_tot} and \ref{fg:likeratiosd_tot}
upper limits on the SagDEG density can be inferred.
In particular, for some WIMP masses and for some halo models,
these limits are comparable or improve the limit already given in ref. \cite{Fre04} 
(of the order of $0.07$ GeV~cm$^{-3}$) on the basis of considerations on $M/L$.

Moreover, figures \ref{fg:likeratiosi_tot} and \ref{fg:likeratiosd_tot} 
suggest that intervals not including $\rho_{sgr} = 0$ at $90\%$ C.L.
exist for some values of $m_W$.
This points out a slightly preference for the presence of a SagDEG contribution in the data.
However, considering the uncertainties on the SagDEG velocity 
and velocity dispersion (that is e.g. superimposing the allowed regions
in the figures \ref{fg:likeratiosi_tot} and \ref{fg:likeratiosd_tot})
in most of the considered scenarios 
the absence of SagDEG is still allowed at $90\%$ C.L..

It is worth to note that
in many of the analyzed configurations the inclusion of the 
SagDEG contribution improves the data fit.
For example, for SI candidate in the case of a 
stream with velocity set $\vec{V}_{sph}$ and velocity
dispersion $v_{0,sgr} = 40$ km/s, 
about $67\%$ of the configurations have an improvement of the
data fit by the inclusion of the 
SagDEG; in particular, the improvement of about 18\% of them is 
better than 2 $\sigma$.

\begin{figure} [ht]
\centering
\includegraphics[width=200pt] {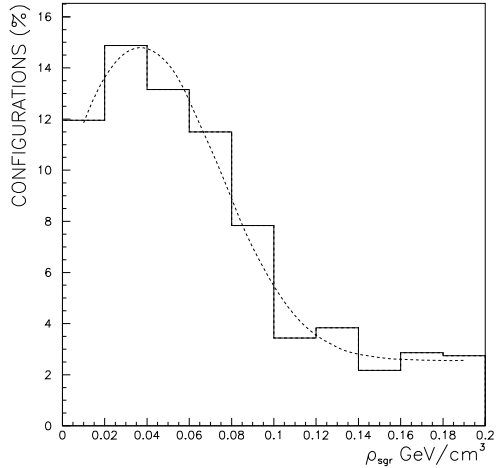}
\caption{Example of the cumulative percentage distribution of $\rho_{sgr}$ 
best-fit values providing a C.L. better than 2 $\sigma$ with the respect to
the absence of SagDEG. A pure SI candidate and fixed SagDEG stream with
velocity set $\vec{V}_{sph}$ and velocity
dispersion $v_{0,sgr} = 40$ km/s have been considered.
About $60\%$ of these $\rho_{sgr}$
best-fit values are below 
$0.1$ GeV cm$^{-3}$. 
See text for implications. 
}
\label{fg:bestrho}
\end{figure}

Another interesting information can be inferred by studying the $\rho_{sgr}$ best-fit values 
achieved for the various considered models.
For this purpose,  
the cumulative percentage distribution of 
$\rho_{sgr}$ best-fit values providing a C.L. better than 2 $\sigma$ with the respect to
the absence of SagDEG
is shown in Fig. \ref{fg:bestrho}. A pure SI candidate and fixed SagDEG stream with
velocity set $\vec{V}_{sph}$ and velocity
dispersion $v_{0,sgr} = 40$ km/s have been considered here as an example.
About $60\%$ of these models gives $\rho_{sgr}$
best-fit values below
$0.1$ GeV cm$^{-3}$; in addition, the distribution 
peaks around $\rho_{sgr} \sim 0.04$ GeV cm$^{-3}$.
These latter values are intriguing considering the expectations on 
the stream density at Sun position -- that is few $\%$ of the local dark halo -- 
based on some theoretical studies about the disruption
of the satellite galaxies falling in the Milky Way halo
\cite{stiff}.

This preliminary analysis offers hints on the possibility to 
investigate halo features by annual modulation signature
already at the level of sensitivity provided by DAMA/NaI.

\section{Conclusion}

In this paper a preliminary study on the effect of the presence of Dark Matter particle streams
in the galactic halo has been analysed on the basis of the annual modulation data collected by DAMA/NaI.

In particular, the case of the Sagittarius Dwarf Elliptical Galaxy (which presently is the better known case)
has been discussed here showing its effect on the allowed volumes/regions 
for some astrophysical, nuclear and particle physics
scenarios related to the case of WIMP candidates. 

The potentiality of a similar approach to 
investigate the halo composition has also been pointed out as well as the possibility
to derive experimental bounds on the possible contribution
of the SagDEG to the local dark matter density. For some of the 
investigated WIMP masses, the order of magnitude of these bounds,
obtained by local measurements 
\footnote{We remind that the local DM density in the solar system is very 
unconstrained; some limit from planets perihelion precession precise measurements
gives for example $\rho_{Solar system} \lsim 10^6 \rho_{halo}$ \cite{precession}.}, 
is in agreement with 
the existing bounds based on non-local M/L ratio observations.

Other candidates and other non thermalized substructures 
will be addressed in near future studies; in particular,
the availability of larger exposures by DAMA/LIBRA will
offer the possibility of more efficient discrimination capability among 
different possible scenarios.

\end{document}